# Understanding the Governance Challenges of Public Libraries Subscribing to Digital Content Distributors

YUNHEE SHIM, Rutgers University, USA
SHAGUN JHAVER, Rutgers University, USA

As popular demand for digital information increases, public libraries are increasingly turning to commercial digital content distribution services to save curation time and costs. These services let libraries subscribe to pre-configured digital content packages that become instantly available wholesale to their patrons. However, these packages often contain content that does not align with the library's curation policy. We conducted interviews with 15 public librarians in the US to examine their experiences with subscribing to digital distribution services. We found that the subscribing libraries face many digital governance challenges, including the sub-par quality of received content, a lack of control in the curation process, and a limited understanding of how distribution services operate. We draw from prior HCI and social media moderation literature to contextualize and examine these challenges. Building upon our findings, we suggest how digital distributors, libraries, and lawmakers could improve digital distribution services in library settings. We offer recommendations for co-constructing a robust digital content curation policy and discuss how librarians' cooperation and well-deployed content moderation mechanisms could help enforce that policy. Our work informs the utility of future content moderation research that bridges the fields of CSCW and library science.

## 1 INTRODUCTION

Public libraries provide users with carefully curated and reliable information on various topics [1-3]. Librarians usually determine the offered content by following their library's collection development policy. Such policies incorporate values essential to the libraries' public service goals, such as community engagement and inclusion [4, 5]. Further, libraries' curation practices seek to protect people from harm while valuing freedom of expression and diversity of thought [6].

In recent decades, and especially after the covid-19 pandemic, public library users' demand for digital content, such as e-books and streaming videos, has proliferated. In response, many public libraries have turned to large-scale digital content distributors, such as Hoopla[1] and OverDrive (Libby).[2] These services provide libraries with vast amounts of preconfigured digital content immediately available to their users. However, such external subscription services mean that libraries outsource their digital offerings' content curation.

---

[1] https://www.hoopladigital.com
[2] https://www.overdrive.com

Doing so raises questions about the quality of offered content, the labor of maintaining content quality, the distribution of responsibility between librarians and distribution service staff, and the extent to which the curated digital content aligns with libraries' public service goals.

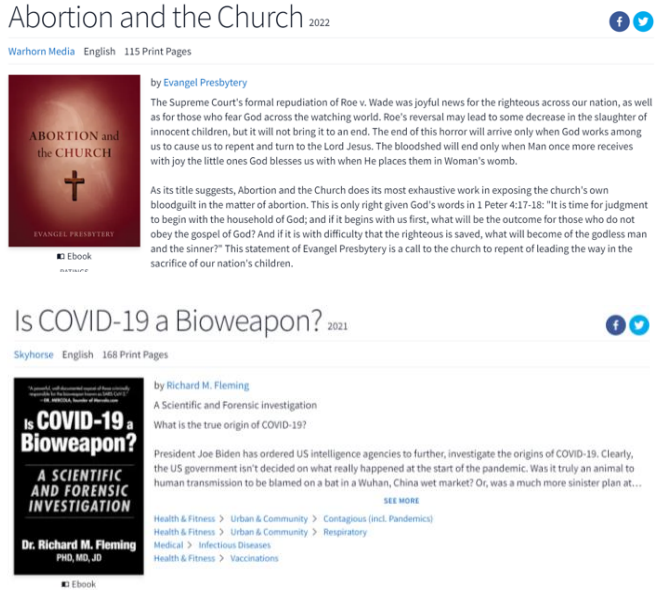

*Figure 1 : Examples of e-books promoting (a) hate against women choosing abortion and (b) misinformation and conspiracy theories about covid-19. As of Jan 9, 2023, these e-books show up among Hoopla's top search results for queries on 'abortion' and 'covid-19'.*

Concerns about public libraries' reliance on digital subscription services are rising due to the growing number of anecdotal examples where inappropriate materials have been found on digital distribution platforms [65]. For example, Hoopla's website shows e-books that include hate speech against women choosing abortion and covid-19 misinformation when searching for the keywords "abortion" and "covid-19," respectively (see Figure 1). Such titles would likely violate most libraries' content curation policies and be excluded from their collections. However, the Hoopla service provides library users access to such titles. Worse, such problematic materials appear among the top search results on the distributor's platform. When this happens, some users may believe these materials are authoritative sources since they are available through trusted libraries.

This digital content curation problem has not escaped the attention of public librarians. Popular librarian groups, such as the Library Freedom Project and Library Futures, have raised concerns about this issue [65]. Several librarians from these groups have voluntarily reviewed the content in digital distribution platform, such as Hoopla, and documented many inappropriate materials, such as those featuring Nazi slogans and anti-LGBTQ+ rhetoric [15, 39]. Librarians have raised alarm bells about such inappropriate materials and highlighted other critical problems with digital distribution services, such as their lack of transparency and accountability about curation practices [48]. They have also called for these services to take urgent steps to address this issue.

Despite calls for change by many librarians, relatively little is presently known about libraries' overall experiences subscribing to digital distribution services. Since libraries still subscribe to them even though distributor services may provide inappropriate content for users, we must first understand their appeal to the libraries. We therefore ask:

*RQ1: What benefits do libraries incur when they subscribe to digital content distribution services?*

Librarians usually curate the physical collections themselves. However, when they rely on the pre-configured package of digital content distributors, they effectively outsource content curation to those distributors. Doing so raises the question of how well the librarians understand these distributors' content curation processes. We sought to explore what librarians think about the transfer of curation control and how they view distribution services' moderation practices. Therefore, we ask:

*RQ2: How do librarians perceive digital content distribution services' content curation and moderation practices?*

Subscribing to digital distribution services may introduce new content governance challenges for libraries, such as identifying inappropriate ebooks. Therefore, we explore how such subscriptions create new tasks for librarians. We ask the following question:

*RQ3: How does subscribing to digital content distribution services affect the work of librarians?*

The current literature on these content curation practices is scarce. Therefore, we turn to the more advanced literature on social media moderation for theoretical guidance on this topic. We observe many similarities between the content curation challenges of digital distributors and social media platforms. Both services host digital content for consumption by the public [4, 7, 8]. Some offered content can be offensive, misinformative, or otherwise inappropriate, leading to harm. Both services contend with the challenges of scale since it can be prohibitively expensive to review all included materials manually. Social media sites have invested in constructing complex content moderation infrastructures and processes to serve their users better [9]. Therefore, we examine how we can apply the learnings from that literature to inform the practices of distribution services working with public libraries. We observe that ensuring rigorous quality control of offered content is even more critical in the public library context than on social media sites because individuals seeking information usually have a higher expectation of reliable information when using libraries [10].

We interviewed 15 public librarians with experience managing or servicing at least one digital distribution service to answer our research questions. We have complemented this interview data with a review of content curation information on distribution service websites and news articles on this topic. Our analysis reveals that public libraries subscribing to pre-configured digital services like Hoopla enjoy several benefits, such as affordable and immediate access for their users to a vast amount of popular digital content. However, these subscriptions introduce challenges, such as increased availability of low-quality digital content and a reduced understanding among librarians of why certain digital materials were included in the service collections. These problems raise concerns that users and funding organizations may reduce their trust in libraries.

Further, we found that low-quality content and a lack of transparency about content curation cause librarians to desire greater engagement with distribution services. However, these services do not offer rich communication options to librarians, and library staff has limited control over curating the content made available through these services. Librarians cannot act ex ante but are limited to regulating inappropriate content ex post [11]. We document how this shift in curation authority changes librarians' work.

Our analysis concludes by suggesting that distributor services should carefully design a content curation policy and make it accessible to all subscribing libraries. This policy should serve libraries' public services goals, such as community engagement, inclusion, and diversity. Further, the distributor services must curate the materials they include in their services in alignment with this policy. We also recommend that distribution services implement more advanced communication between their staff and librarians to understand libraries' evolving needs better and update their practices to respond to them. Our findings identify opportunities for distribution services to use the professional expertise of librarians to help them design curation policies and moderate inappropriate content. Further, we discuss the role of AI-based moderation techniques in addressing the challenges of scale and reducing the need for human labor. We also examine how other stakeholders, such as library associations, users, and researchers, can improve current curation and moderation practices and continue garnering public trust in libraries.

## 2  BACKGROUND AND RELATED WORK

This section first offers background information on digital content curation in public libraries and identifies the critical gaps in our understanding of the digital subscription services. Next, we review elements of social media moderation research that can usefully inform distribution services' governance approach.

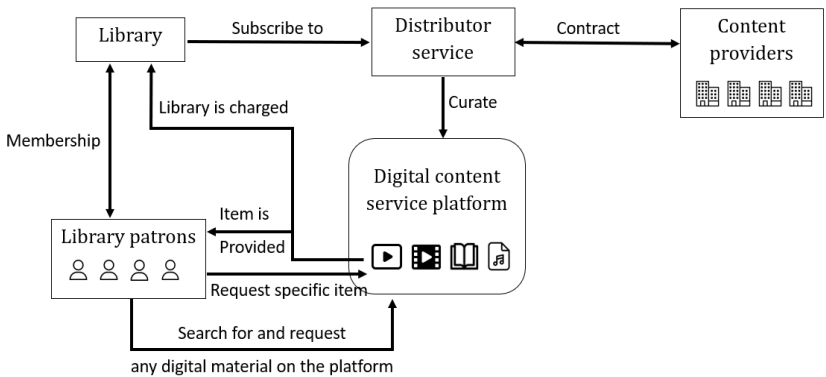

Figure 2. Pay-per-view model (used by Hoopla): Distribution services contract with content providers to curate digital content for their platform. When a library subscribes to the service, its users can access all content offered by that service. The library only pays the service for the content that user accessed.

### 2.1   Digital Content Curation in Public Libraries

Public libraries have served users with digital content such as e-books since the mid-1990s [12]. However, with the popularity of digital content, libraries now serve many digital materials with the help of content distributors such as Overdrive and Hoopla. Though librarians still curate physical collections themselves, distributors act as wholesalers for digital content, managing the negotiation, licensing, and billing of digital materials with their publishers on subscribing libraries' behalf. They also offer an online platform where library users can use, search for, and access the available content [13]. Larger libraries let users use their library website's search function to show all content available through the digital distribution services that they have subscribed to. Selecting a digital item in these search results redirects users to its corresponding page on the distribution service website to access that item. In contrast, many smaller libraries do not populate their website's search databases with subscribed services' content; they ask users to visit the distribution service website directly to search for and access the desired content.

Content distributors vary in their pricing mechanisms. For example, Hoopla offers a *pay-per-view* model (Figure 2), in which libraries make all the content curated by the distributor available to their patrons and pay the distributor only for the content their patrons use. In contrast, Overdrive adopts a *selection* model in which distributors curate the digital content to configure a package; librarians select to pay for access to specific items in the package, and patrons can use only the paid-for items.

In this research, we focus on the pay-per-view subscription model, which offers lesser curation control to the subscribing libraries than the selection model. Many libraries adopt this model because of its advantage of incurring relatively low costs to obtain each title [14]. Prior research on electronic content in libraries has shown how library administration can cost-effectively subscribe to digital content services [12, 13, 15, 16]. We explore here the other benefits that this subscription model offers to libraries.

While many public libraries subscribe to pay-per-view distribution services, it is unclear how the library staff perceives and responds to this subscription model. This paper takes the first steps toward addressing that gap. By highlighting librarians' perspectives, we surface the challenges they face in satisfying their patrons' digital content needs with this model. Prior research has documented libraries' quality control challenges when working with large sets of purchased e-books [17]. We build upon it to examine their quality control challenges when digital content is accessed through a subscription rather than a purchase. By examining how the introduction of commercial digital distribution services influences library databases, we also contribute to critical information studies [18-20].

Public libraries have conventionally offered reliable and informative physical collections on various topics. Librarians oversee collection building, including selecting and purchasing relevant content, by carefully evaluating whether each content candidate fits the collection in question [21]. This curation is made in compliance with the library's collection management criteria [17]. However, libraries outsource this collection curation to digital distributors when opting for the pay-per-view digital subscription model [22]. An inevitable consequence of relying on digital distributors is the introduction of content in library collections that does not align with the library's curation criteria [16]. Outsourcing also reduces the level of control libraries exert in managing their offerings. We examine this change and its consequences in this research.

Another consequence of relying on the subscription model is that it obligates library staff to conduct an additional evaluation of the subscribed content to provide reliable and unbiased information to the public [22, 23]. This evaluation increases the content management tasks that librarians must execute. Otherwise, they risk losing public trust when they fail to identify inappropriate or outdated content before users access it [24]. Therefore, how digital distributors curate and manage their subscription packages directly affects the workload of public librarians. We examine the changes in librarians' work when subscribing to distribution services in this research.

## 2.2   Content Moderation

Grimmelmann defines *content moderation* as "the governance mechanisms that structure participation in a community to facilitate cooperation and prevent abuse" [11]. Content moderation, a critical offering that social media platforms provide, is integral to running online communities successfully [25] and ensuring the safety and security of users [7]. Some key aspects of content moderation include deprioritizing or removing content that violates community norms and sanctioning users who post such content [26-28]. When done well, moderation actions can improve community outcomes by rendering the most relevant content more visible, encouraging users' compliance with community norms, and building a group identity based on the shared values of community members [29-31].

Prior HCI (Human Computer Interaction) and CSCW (Computer Supported Cooperative Work) research has characterized the content moderation policies across social media sites [7, 30, 32-34], examined the role of automation and human labor in enacting content moderation [26, 35-38], and documented the need for and effects of incorporating transparency in moderation decisions [39-42]. We build upon this literature to examine analogous concepts of curation policies, distribution of labor, and curation transparency of distribution services serving libraries.

Recently, researchers have begun expanding the content moderation literature beyond social media. For example, McKinnis et al. examined how news companies, like *The New York Times*, manage online discussions on their websites [43]. Following this tradition, we extend content moderation research to examine how digital distribution services serve public libraries. By doing so, we aim to start a conversation between the content moderation researchers in the CSCW and library science communities since their content management challenges are phenomenologically alike. The theoretical and empirical insights from each community can inform the other.

Content moderation includes curating a set of community policies or guidelines that shape user activities within each community [25]. These policies can be either prescriptive or restrictive. *Prescriptive* policies explicitly specify the recommended behaviors in an online community. *Restrictive* policies limit behaviors that are not welcome in a community [33]. Social media moderators enforce the latter by sanctioning the content or accounts that violate them. Traditionally, libraries' curation policy focuses on what content to include rather than exclude. However, the new digital subscription model has complicated librarians' content curation practices, where they must now also attend to what should be restricted. We explore librarians' shift toward restrictive actions in this research.

Prior examinations of policies across social media platforms have found that some of the most frequently stated policy goals address hate speech, death, and rape threats [44], as well as other antisocial behaviors, such as trolling, online harassment, and virtual rape [31]. Such actions disproportionately affect marginalized populations [45, 46]. Therefore, the policy focus on preventing them suggests that social media platforms recognize the need to protect vulnerable groups. Similarly, digital content that incorporates racist or sexist biases would disproportionately affect marginalized groups. Therefore, we must consider whether digital content subscription services and curation policies create unequal conditions or promote unjust outcomes [4, 7, 8]. Our research explores how we can advance equity and inclusion in the digitally offered library content and how social media moderation policies can inform this goal.

As social media platforms continue to increase the number of users they serve, they have had to scale up their moderation infrastructure to regulate inappropriate content and keep their sites usable and efficient. With the challenges of scale that attend massive digital distribution services like Hoopla, large-scale social media moderation could provide valuable guidelines for scale management. Content moderation decisions on social media are made by both human moderators and AI-based moderators (or auto-moderators). Auto-moderation reduces exposure to violent or offending content for human moderators, offloads the content that requires human review, and facilitates consistent and efficient moderation [38].

Recognizing these benefits, several studies have attempted to improve AI-based moderation systems [26, 27, 35]. However, human moderators remain indispensable. For example, volunteer moderators of multi-community platforms like Reddit and Facebook contribute their ingenuity, imagination, and dynamic human insight to evolve rules based on their community's changing needs [31]. They can make such rule changes meaningfully democratic and enact nuanced moderation decisions in many cases that are difficult for automated systems to adjudicate [26]. Through these mechanisms, human moderators help construct a sustainable environment for the community they moderate [47] and shape its identity. We explore the division of labor between human and automated review that could shape digital content curation and identify additional challenges this division may face when moderating digital content for libraries.

Social media scholars have noted that content moderation is influenced by and affects various stakeholders, such as platform users, commercial and volunteer moderators, platform employees and owners, corporate boards, regulatory agencies, legislators, activists, academics, and journalists [7, 8]. Similarly, the governance of libraries' digital content is also relevant to many stakeholders, such as library patrons, librarians, funding organizations, publishers, distribution service employees, library consortium[3] staff, library associations,[4] and lawmakers. We consider the multi-stakeholder nature of the digital subscription model and examine the role that some key stakeholders play and could play in digital content moderation.

A critical aspect of content moderation is transparency about its procedures and outcomes. DeNardis and Hackl [48] highlighted that transparency directly affects user participation on social media [48]. Specifically, social media users consider content moderation fair when they perceive that community rules are clear and easy to follow and receive feedback about rationales for their post removals. Such perceived fairness, in turn, positively influences participation and norm compliance in online communities [39, 40]. When community rules are explicit, newcomers understand what is expected of them and more efficiently assimilate into the community [49]. Inspired by these insights, we examine the role that transparency in digital content moderation plays in library settings.

## 3   METHODS

To examine the nuances and consequences of digital content curation through digital content distributors, we conducted semi-structured interviews with public librarians who work at libraries that subscribe to large-scale digital distributors such as Hoopla and OverDrive. We contacted customer services, sales representatives, and content specialists at Hoopla and OverDrive to understand their perspectives on content curation, but we did not receive any response to our interview requests. Therefore, we have relied on interviews with librarians as our primary data source and provided an in-depth analysis of the librarians' perspectives. We have complemented this by reviewing content curation information available on the distribution service websites and news articles about their curation operations to enhance further our understanding of these services' content management practices. The Rutgers University IRB approved this study on 22 September 2022.

---

[3] A library consortium is a joint group of libraries that share the library system for patrons and purchase the offered content as one collective, central body. Member libraries can provide a larger volume of content to their patrons by effectively pooling their budgets to buy shared materials.

[4] Library associations are nonprofit organizations that promote the interests of people who rely on libraries and information professionals. Examples include the American Library Association (ALA) and the International Federation of Library Associations and Institutions (IFLA).

## 3.1 Participants

We interviewed 15 public librarians with experience working in a library setting with a digital distribution service subscription. We limited our participant sample to public librarians because public libraries serve patrons from all walks of life, including vulnerable individuals such as children, immigrants, and the homeless [50]. The diversity of individuals they serve also informs the practices of public libraries, with their focus on providing balanced, accurate, democracy-supporting information. Our focus on public libraries helped us explore how librarians seek to uphold these values when working with digital distribution services.

We continued to recruit interviewees and collect data until our analysis reached theoretical saturation. All participants are public librarians with positions in digital content administration. Table 1 lists their demographic details and official job titles. They are all White and have received a Master's degree education. Of the 14 libraries our participants are affiliated with, 13 currently have Overdrive subscriptions, and 12 currently have Hoopla subscriptions, although most participants were familiar with both the subscription models.

Table 1. Demographic and Employment Information of Study Participants. Unavailable information is indicated as "-."

| No. | Age | Gender | Job Title | Experience (in years) |
|---|---|---|---|---|
| **P1** | 40 | Male | Head of Information Services | 15 |
| **P2** | 61 | Male | Library Director | 37 |
| **P3** | 44 | Female | Digital Resource Director | 19 |
| **P4** | 48 | Male | Consortium Director | - |
| **P5** | 53 | Female | Library Director | 26 |
| **P6** | 39 | Female | Head of Reference and Adult Services | 15 |
| **P7** | 42 | Female | Outreach and Diversity Librarian | 16 |
| **P8** | 54 | Male | Head of Adult Services | 32 |
| **P9** | 39 | Female | Adult Services and Programs Librarian | - |
| **P10** | 60 | Female | Head of Information Services | 35 |
| **P11** | 39 | Male | Library Director | 14 |
| **P12** | 58 | Female | Digital Resource Director | 20 |
| **P13** | 48 | Female | Library Director | 16 |
| **P14** | 56 | Female | Digital Buying Lead | 33 |
| **P15** | 35 | Female | Collection Development Specialist | 11 |

## 3.2 Data Collection

To recruit participants for this study, we first contacted a librarian at our institution who connected us with a New Jersey-based consortium of public librarians in her professional network. We requested staff members at this consortium to disseminate our recruitment message to relevant librarians. To get more diverse perspectives, we similarly contacted consortiums based in New York, Boston, and Philadelphia. These consortiums assisted us by distributing our recruitment message on their library community listservs and message boards. They also connected us with librarians who could serve as interview participants. Additionally, we used snowball sampling after the interviews began to elicit recommendations from some participants about who else we could interview. We conducted all interviews between October 6th and December 7th, 2022.

The interviews lasted between 52 and 92 minutes each and were conducted over Zoom, an online conferencing tool. We recorded all interviews with the participants' consent. Our semi-structured interviews consisted of three phases: first, we asked participants which digital distribution services their libraries subscribe to, the librarians' involvement in digital content curation, and the details of communication between librarians and vendors. Second, we asked questions about their perspectives on the content curation of distribution services and the benefits, deficiencies, and challenges that the subscription model presents. Finally, we asked participants about their collection management policies and how digital distributors can modify their policies and practices to address the prevalent problems.

As the interviews progressed, we revised our interview guide to gain deeper insights into some emerging concepts. Participants freely shared their experiences and ideas and were allowed to skip questions and reflect more on any issues that arose in the discussions before the end of the interview.

### 3.3  Data Analysis

All interviews were conducted in English and transcribed. Next, we applied interpretive qualitative analysis to our transcripts [50]. First, we uploaded all interview transcripts to NVivo 12 software for qualitative coding. We then read each transcript multiple times to familiarize ourselves with the interview data. After this, we conducted open coding on a line-by-line basis to identify emerging concepts and stay close to the data. We then performed multiple iterative rounds of coding and memo writing.

Through this process, we began to group the preliminarily found child codes into parent codes that all pertained to various salient issues in the digital subscription model. Our memo writing helped us describe emerging themes and deepen our reflection on the relationships among them. The authors discussed the evolving codes and memos throughout our analysis. Once our analysis began converging, we followed an iterative process of dividing and aggregating various codes using sticky notes, which yielded 11 parent codes and 45 child codes. We combined and distilled these codes into key themes, which we present next as our findings.

## 4  FINDINGS

### 4.1  Benefits of Subscribing to Digital Content Distributors

Libraries subscribe to digital content distributors like Hoopla and OverDrive because they present several benefits. First, these distributors offer a large volume of content to the libraries, and participants considered these subscriptions more affordable than buying unpackaged digital content. Unlike physical books, electronic content takes up no library shelf space. The distributor's servers also host this content, so the libraries need not devote server space to store it. Therefore, subscribing libraries do not need to limit the content volume the distributors provide. As one participant observed:

> *"It's just very difficult to replicate that [volume] in any other way. .... I think that is what drives the popularity of Hoopla in our library and other libraries as well."-P6*

Second, these distributors' packages offer a broad spectrum of content types. They provide e-books, audiobooks, movies, TV shows, magazines, and music that library patrons may borrow:

> *"Hoopla is offering different things like TV shows or more recent music releases, and then we can just have it in the catalog and pay for it when they [patrons] choose to borrow it." - P11*

A third benefit of subscribing to these distributors is that many materials they offer are popular among library patrons. Finally, the provided content is immediately available to the library patrons on their computers, phones, tablets, and TV at any time they desire:

> *"Well, the strength of Hoopla is definitely that all of it was simultaneously usable." -P15*

Participants especially noted the benefits of the pay-per-view subscription model in improving access to digital content:

> *"Hoopla, the items are always available; you don't have to wait on hold for them. And there's also not the model that Libby (OverDrive service) uses where you get 25 checkouts, and then you have to purchase it again." - P7*

### 4.2  Distributor Packages Contain Inappropriate Content

Our analysis shows that libraries subscribing to digital distribution services face quality-control challenges. Several participants described finding content in the subscription packages that they felt was unfit for their patrons' needs:

> *"They just provide access to everything, which is what they advertise as their attraction, which also means there's a lot of junk." - P3*

Another participant presented examples of inappropriate content he found in the distributor package:

> *"They have some materials on there, and you don't initially know what it is until you do some digging. Um, but if you begin digging into that collection a little bit, you realize there are some kinds of pornographic and obscene materials in that collection." - P2*

Such inappropriate content risks harming patrons who encounter it, e.g., minors may get exposed to violent pornography. Our participants were especially apprehensive about patrons' access to misinformation. One critical mission of public libraries is to deliver reliable and accurate information to their patrons. However, our participants noted that achieving this mission becomes challenging when using distributor packages because they often contain outdated content or misinformation:

> *"I think Hoopla could be a lot more selective about what is in the collection because I don't think that they pay any attention to how old titles are and whether the information is outdated." - P3*

Participants noted that their concerns about misinformation have only increased in recent years. Given the rise of inaccurate information after the covid-19 pandemic began, providing credible health information has become a critical goal of many public libraries [51]. However, distribution services offering misleading health-related content undermine this goal. For instance, a participant described examples of such content that she found in the distributor service's platform:

> *"Things that just are outright false about vaccines, such that vaccines cause autism or covid was planned. There's just a lot of stuff that's outright racist...I think the largest issues were racism and inaccurate things about health." - P7*

Similarly, another participant worried about the negative consequences of having e-books with health misinformation for library patrons:

> *"We found a lot of problematic medical covid books that were put out by either non-medical professionals or people who, once we began to research them, had lost their medical licenses, and been discredited. But the books were being cataloged in Hoopla as medical information." - P6*

### 4.2.1   Consequences of libraries hosting inappropriate digital content

Participants maintained that the problem of inappropriate content in digital distribution services poses severe consequences. They felt that this problem would discredit entire library services if it remained unaddressed. Participants mentioned that patrons encountering inappropriate content would perceive it as the library's selection failure, undermining their trust in libraries. Over the long term, such failures could also affect how organizations that support libraries view them.

> *"Say, they're putting something antisemitic in without me knowing, and my user finds out – that causes so many problems for us. Not even on a local level, on a trust level with the library. How could you do business with a company that has this material? It could hurt future fundraising and development." – P13*

### 4.2.2   Inappropriate digital library content is more challenging to detect than physical content

As noted, participants detected the prevalence of low-quality digital content, such as pornographic information and misinformation, in their libraries made available through distribution services. However, they reported receiving fewer patron complaints about such content than they did for physical collections. Instead, patron requests for digital content usually relate to buying new content they need. This difference makes it harder to detect inappropriate digital content than physical content. One participant pointed out that the probability of patrons finding harmful digital content is lower because they usually encounter it through an online search, and they may not use search keywords that show results containing harmful content:

> *"I think e-books are a little different than print books because when somebody's in a library with a print book right in front of them, walking down an aisle, they see the book, whereas the [digital] titles, unless you're looking for them, are actually kind of hidden online." – P10*

The individualized nature of search makes some digital content much more discoverable than others. While this reduces the probability of any inappropriate digital material inflicting harm, it also makes it more likely that it remains available through the library. In contrast, all physical content in libraries is usually laid on library shelves where users can quickly scan them. Thus, it is more likely that a patron would encounter objectionable physical content serendipitously and report it to the library staff.

*4.2.3  A few distribution services have a monopoly in the marketplace*

The problem of content distributors hosting inappropriate content is especially damaging when we consider that librarians have little choice regarding digital content providers. Our participants mentioned that most libraries subscribe to one of a few large-scale distribution services, such as Hoopla or OverDrive. By offering an extensive package of electronic content, these services place themselves in a unique position in the marketplace compared to competitors. Participants P7 and P9 told us that few other digital distributors could compete with these large-scale content providers.

> *"So, unfortunately, there seems to be a little monopoly on e-book content and audiobook content. There are really not that many other models or companies to go to." – P7*

> *"I think they are pretty prominent right now. I don't see them going away anytime soon. OverDrive itself has been swallowing up so many little ones, like Universal class, and Kanopy." -P9*

Our participants worried that this marketplace monopoly creates a power dynamic between distribution services and libraries, where the former has almost complete autonomy in structuring their service models without any oversight. This leads to situations where librarian concerns may remain unresolved.

**4.3   Concerns about the Curation Policy of Distributor Services**

Our analysis suggests that transparency about content curation policy is an essential concern for librarians. All our participants told us that the library they work for had established a content curation policy for their physical collections. Traditionally, librarians have identified the needs and interests of their communities and incorporated them into content curation through forming a 'collection management' policy [21]. Moreover, public libraries usually post this policy in a physical place or on their website to inform users how they curate their collections. This measure illustrates how libraries prioritize and implement policy transparency as service providers to their users.

> *"We have a collection development policy, which I can share with you. I think it's on our website." – P6*

Outsourcing digital content curation to distribution services makes it difficult for librarians to fully know how the content is curated or which curation policies and criteria shape the digital collections. Our participants observed that low-quality, even harmful, content in the subscription service packages suggests their poor content curation practices. Some participants wondered whether these services even have a curation protocol in place:

> *"For Hoopla? I think they accept everything and anything. I honestly do believe that there are no stipulations. I think, whatever is submitted to them as a published title, they take." – P7*

Participants mentioned that their understanding of how the distribution services curate their content suffers because these services do not provide sufficient details about their curation policy. For example, Hoopla offers little information about its digital content curation on its website, and that information was last updated in 2015 [52]. The webpage containing this information notes that Hoopla's *"selections are made to provide depth and diversity of viewpoints to the existing media collection,"* but it does not reveal any implementation details of how it achieves this goal.

> *"As far as I know, they don't really reveal anything to us about how they're collecting. They build the content for their collections – I don't know." - P2*

Participant P3 argued that the distribution services should publish their content curation policy and methods and communicate them to the library staff, especially those who directly face the patrons. For example, when she found e-books expressing Holocaust denial in the service package, she expected the service would proactively detail the reason for including such content:

> *"I meant to stay on top of that after the issue with the Holocaust-denying books. But if they have responded to that by providing a policy about their selection, I haven't seen it yet. So, they definitely didn't proactively send it out to me." - P3*

Suspecting that the digital distribution services do not have any curation policy, a participant suggested that libraries can assist content distributors with policy creation by sharing their library's carefully designed policies for content curation and moderation.

> *"We sent them our collection development policy, which is obviously several pages long and recently revised. And we made it clear these are the standards we're holding ourselves to. This is why we're holding your collection to a similar standard because, on the user end, your collection is our collection, and vice versa." - P6*

### 4.4   Concerns about How Distributor Services Handle Inappropriate Content

Our participants mentioned that their library also has protocols for removing inappropriate content in addition to a content curation policy. Library patrons can request that inappropriate content they come across be removed. There are processes to review such feedback and take action to enhance the patron experience. These processes are called 'reconsideration requests' or 'book challenges.' One participant noted:

> *"At the library, we have a suggestion box. We also have an open, like, suggestions email that's just for whatever users want to comment on … We do have our policy about filing withdrawal requests for objectionable material." - P11*

Participants mentioned that reviewing patron requests for content removals could be complex. The librarians first investigate the content and make decisions based on their policy. However, in some problematic cases, the content review is escalated for further discussion with an executive board or a review committee before making a final decision:

> *"I remember a movie was challenged, and the director put together a committee or a task force to look into it. We had to actually watch the movie." - P12*

Once librarians decide on content review, they do their best to explain their reasoning to their patrons. Participants felt that such efforts help users better understand how their library curates its contents. One participant described librarians' broader efforts to educate users about the library's curation philosophy:

> *"If someone complains about a book, we talk to the people first and try and explain to them that we carry books on all different viewpoints." - P14*

Participants expected digital distribution services to have similarly rigorous content review and explanation processes. For example, Participant P4 pointed out the need for quality control of digital content in distribution packages, especially because they include content from publishers and sources that are not considered authoritative materials on the topics they cover. Similarly, another participant observed:

> *"[(The distribution service)] should always think of that as, what happens when things are not good, what is the plan for that? And do you have a policy in place to deal with that?"- P11*

In contrast to libraries' sophisticated and closely monitored 'reconsideration process' for physical collections, the distribution services offer rudimentary support pages that our participants deem insufficient to satisfy users' content review needs:

> *"I'm not seeing anything called out on their [distributor's] help page for objections. It just says, 'contact support,' and when you click on that, it says, 'Need more help? Access the Help form.'." - P11*

*4.4.1 Free speech concerns about moderating digital content*

When moving to the digital subscription model, the content curation focus shifts from what is *included* to what is *excluded*. This shift to excluding materials brought into sharp focus for our participants how their content curation activities may raise free speech concerns. For instance, some participants mentioned that in making content removal decisions, they are concerned about infringing on the freedom of speech of the corresponding content creators included by the distribution services. Additionally, they were sensitive to not infringe upon other users' right to read when removing digital content even when some users may find that content controversial.

> *"We wouldn't remove it [if users challenge the content] anyway because there's no compulsion for you to borrow it. Somebody else, I'm sure, might be interested. ... We support intellectual freedom and the freedom to read."-P1.*

However, many participants acknowledged their obligation to curate the best possible content for each topic; they argued that this curation does not necessarily imply free speech violations as the focus is on offering the best available materials:

> *"We understand that everyone has freedom of speech. However, just because it's on that platform doesn't mean we need to select it for our collection." -P7*

Therefore, like social media governance [53], librarians' concerns show that determining the fine line between respecting free speech concerns and protecting users from harm is a significant digital content governance challenge.

### 4.5 Librarians Hold Little Authority but Incur Responsibility for Digital Curation Decisions

Libraries that subscribe to the pay-per-view model receive the distribution service's pre-configured package wholesale. Within this model, librarians cannot preselect what to include in the package. This lack of authority prevents librarians from adjusting the digital content offered by the package to ensure compliance with their collection management policy.

> *"We don't curate for Hoopla. That's curated for us from the vendor." - P12*

> *"Hoopla is more of, they just put everything up there, and we have no power in the collection development process." - P7*

While the distribution service unilaterally curates its package, library patrons are usually unaware of its role. Therefore, they hold the library staff, who are on the front line of servicing digital content, responsible for curation deficiencies. This burden of responsibility in the face of a lack of authority to select the offered content dissatisfies librarians:

> *"The users have no sense that we didn't buy those titles. A lot of users don't even realize that the vendor is a vendor. They just see the [library name]. And to them, Hoopla and Libby are us. So, they come to us for any questions that they have." - P6*

Some elements of the distribution service's webpage interface further bolster the perception that the library staff is responsible for the content. For instance, Participant P9 noted that the distribution service platform induces patrons to ask the library staff for help by providing the library's contact information. As a result, all patron problems related to the digital distribution service are primarily transferred to the library. Additionally, the distribution services offer no means for patrons to directly communicate with them, obligating libraries to address any challenges patrons face.

> *"[Content providers do] not [conduct] as much one-on-one interaction with the library customer. And virtually no interaction with the user customer." -P11*

### 4.6   Concerns about the Labor of Moderating Digital Content

Library subscriptions to digital content distributors create additional work for library staff. Some participants mentioned that they audit the digital content offered by the distributor platform to ensure that their patrons do not encounter any inappropriate content:

> *"Hoopla, they don't consult us on what they buy every month. They release a whole bunch of new content. So, what I have done is I've made a calendar reminder to myself to look once a month to see what they've added in the last 30 days to see what we can focus people's attention on." - P5*

Participants mentioned frequently auditing digital content by evaluating the publishers' reputations. They make a list of low-quality publishers and send it to distributors, asking them to block those publishers' content.

> *"We have certain publishers we've asked them [distribution services] to block. We have certain titles that we've asked them to block. My staff specifically went through a publisher list of, I think it was 2,800, publishers and looked up each one to see if they were a reputable publisher." - P6*

One participant characterized such new moderation tasks as 'reverse curation.' She felt this ex post evaluation work is a more significant burden than routine ex ante curation tasks for librarians.

> *"So, I definitely feel that regardless of adding things, choosing what's in my collection is simpler than having to fish around for things I might want to suppress." - P15*

Some participants were overwhelmed by the labor involved in evaluating digital content. For example, one detailed the difficulty of squeezing moderation work into her regular tasks, admitting that this work often remains undone as a result:

> *"I haven't, for instance, gone in there and searched to see if there are any horrible books by people who are denying that the Holocaust happened or anything. It's just not with this job. Having so many things going on at once, I don't go looking for trouble until it finds me." - P5*

Some participants felt that content distributors must shoulder the burden of moderation work, not librarians. For example, one participant suggested that distributors hire staff to carefully monitor digital content to ensure quality control before sending the content to the libraries.

> *"The other way is employing librarians and perhaps editors and quality control staff in their company, on their end, to ensure that the materials in their databases are similar to what libraries would purchase." - P6*

Another participant recommended that distribution services assume moderation tasks but also let subscribing libraries influence how the service executes these tasks:

> *"I think it would be good if they [distributors] had like a steering committee made up of librarians, and obviously, they can't open it up to all of their customers, but it would be good if libraries had a voice in the process." -P3*

### 4.7 Cooperation among Librarians may Improve the Efficiency of Moderation Work

Several participants suggested ways to reduce the moderation labor for individual librarians. For example, one suggested that distributors need mechanisms for sharing each librarian's evaluation outcome for digital content with other librarians who subscribe to the same service.

> *"It would be great if … a company like Overdrive or Bibliotheca could enable librarians to put stars on different books, um, so librarians could see what other libraries thought." - P2*

Such sharing of evaluation outcomes could reduce the duplicate effort of staff from different libraries reviewing and removing the same content independently.

Some participants noted that many smaller, stand-alone libraries have only a few librarians who simultaneously perform numerous tasks, from facility maintenance to overseeing reference services. Such libraries would significantly benefit from sharing digital content moderation task outcomes. Our analysis suggests that such sharing mechanisms would be popular among librarians because participants expressed a desire to engage with other librarians when making moderation decisions on digital content:

> *"If I came across anything really awful on Hoopla, I would probably consult with other librarians in my consortium. Should we ask them [content distributors] to remove this? Is this spreading false information? You have to be really careful in situations like that." -P5*

## 5 DISCUSSION

### 5.1 Designing a Content Curation Policy

Our analysis suggests that distribution services should have clear policies and guidelines that dictate their curation practices. Unlike public libraries, which construct publicly available policies to shape the curation of their collections, content distributors do not reveal their curation criteria (Section 4.3). Many participants wondered whether these distributors even have a curation policy guiding their collection construction (Section 4.3). Well-designed and implemented distributor curation policies could reduce the problem of libraries and their patrons encountering inappropriate digital content.

Currently, a lack of knowledge about distributors' policies forces librarians to double-check the digital content offered by content distributors (Section 4.6). Further, when library patrons uncover inappropriate digital content, the apparent absence of a curation policy inhibits librarians' ability to explain its inclusion in the library collection, diminishing users' trust in the library (Section 4.2.1). Therefore, distribution services must prioritize sharing a sensible content curation policy.

In case they do not already exist, constructing specific content curation and removal policies would require substantial effort from content distributors. Our analysis shows that librarians desire to assist distributors in this process (Section 4.3). Libraries already develop policies for curating physical collections in their libraries. Their experience could, therefore, serve as an asset to structuring the content distributors' curation policy. Though significant differences may exist between the policy needs of digital content distributors and libraries, distributors could reflect on and borrow from the curation policy previously built and revised by libraries or get professional librarians' assistance.

Distributors' content curation policies should be clear and sufficiently descriptive for library staff and patrons to comprehend policy details and how they shape curation actions [39]. Such policies should incorporate the values of public libraries, such as inclusion and community engagement, in guiding what content is included. Social media policy guidelines often seek to protect vulnerable populations, e.g., their policies explicitly prohibit online harassment, a problem that disproportionately affects marginalized groups [45, 54]. Digital distributors' content curation policies should similarly seek to protect the most vulnerable groups, e.g., by excluding digital content that may promote discrimination against them. Further, prior literature on social media moderation policy shows that these policies usually need frequent revision based on changing user behaviors [33, 34]. Similarly, distributors' curation policy should be continually revised in line with the evolving needs and goals of the subscribing libraries, patrons, and the broader society.

**5.2 Enacting Content Regulation to Remove Inappropriate Content**

Our analysis shows that it is vital for content distributors not just to construct good curation policies but also to implement them well and prevent the inclusion of inappropriate content in digital collections. Borrowing from the social media moderation literature [41], we discuss how human and automated approaches to moderation could help flag and remove inappropriate digital library materials.

We propose that professional staff from a content distributor or librarians who take charge of a digital collection could regulate the already included content in the distributor packages. Our participants suggest that digital distributors hire skilled staff to handle the moderation tasks (Section. 4.6). Further, our analysis and prior research suggest that librarians could bring their content curation expertise to evaluate the digital content [6]. Library committees could also evaluate physical collections and make collective decisions on the inclusion of specific materials [38]. However, it is vital that distribution services do not burden the library staff with such labor without compensation. If these services remain resistant to performing moderation tasks on their end, libraries should account for them when making subscription decisions. Further, they could better prepare their workforce for such tasks by recognizing them as part of librarians' duties or hiring professional staff to perform them.

Our analysis shows that mechanisms for libraries to share their evaluations of digital content could significantly reduce the required moderation labor (Section 4.7). We suggest creating online spaces where libraries subscribing to the same digital distribution service could share content that they find inappropriate. This process would be more streamlined if the distribution services themselves created such spaces. Our findings suggest that many librarians would welcome an opportunity to share and receive such information. Many libraries already benefit from the cooperative development and sharing of physical collections through their participation in library consortiums, e.g., by having reduced costs and improved professionalism [21]. Similar cooperation could benefit the co-curation of digital materials.

As Jhaver et al. [39] show, human moderators of social media content can be overwhelmed by the physical and emotional labor this task entails. Asking public librarians to review the distribution service's content could introduce similar challenges. Many interview participants mentioned the need for more time and resources to audit digital content (Section 4.6). Social media systems use automated moderation to address similar challenges of scale [39]. Therefore, we consider whether and how automation could play a role in the library context.

As far as we know, no automation currently exists to assist librarians in auditing the digital content from distributor services. Participants mentioned that they pay attention to metadata, such as the publishers and authors when auditing these digital materials. Thus, constructing training data that captures source metadata and regulation decisions of librarian-audited digital content and building machine learning models on such data could help automate moderation tasks for future content. Extant research on designing innovative human-machine collaboration systems for social media moderation could offer additional guidance on how automation can improve curation outcomes [35, 55]. For example, such systems could include automated mechanisms that provide information cues for librarians to make faster and more accurate decisions.

Leveraging the collective labor of librarians as moderators could also contribute to building a more accurate automatic system for detecting harmful content [45]. Additionally, distributor services could develop mechanisms for library patrons to rate materials they borrow on their appropriateness along several dimensions, such as relevance, offensiveness, and usability (Section 4.4), and share these ratings with the libraries. Such patron ratings could add valuable information to the moderation system by notifying the library staff when a specific material is frequently rated as inappropriate. Evaluating broader patterns in patron ratings could also inform revisions in the library's content curation policy.

Our analysis shows that patrons often incorrectly blame their library staff for inappropriate digital content (Section 4.5). To address this, libraries could put their own online disclaimers on digital collections *not* curated or reviewed by the library staff. They could guide patrons on steps to undertake if they find any digital item that violates the library's curation policy. Libraries could also hold knowledge sessions that educate patrons about digital subscription models and libraries' limitations in reviewing all available digital content. Such actions should empower patrons to become more vigilant knowledge consumers.

Finally, in addition to removing inappropriate content, distributor services and libraries must attend to the order in which different materials are shown to the patrons on their websites. Ensuring that neutral, more popular search requests show diverse and authoritative materials would help reduce access to low-quality content and better serve the patrons.

### 5.3   Incorporating Transparency in Content Curation and Moderation

Our findings reveal that transparency about the governance decisions of digital content is crucially important to librarians. Currently, librarians often do not understand why specific inappropriate materials are included in the distribution package because the distribution services do not provide their reasoning or reveal their curation policies (Section 4.3). Such information would empower librarians to explain the curation reasons to their patrons. Transparency is closely related to accountability for the service's decisions [42], which could be a reason for the current lack of transparency in service operations. However, a lack of transparency creates additional burdens and greater dissatisfaction for librarians and their patrons.

Social media moderation scholars argue that content governance should occur with the cooperation of all key stakeholders [7, 8].In line with this, one way to add transparency to the curation policy and the moderation process is through distribution services having conversations with the library staff and their users. Creating opportunities for communication should, therefore, be an urgent priority. Services currently offer limited means, such as email or phone contact, for individual librarians to communicate with them. Additional town hall discussions or group meetings between the distribution service and the library staff about digital curation and moderation would let the service providers describe their curation processes and understand librarians' needs. These richer ways of communication could also help the distribution services develop or reconstruct their policies to align with the rapidly changing library environment.

Our findings show that librarians often conduct additional evaluation work to verify the content quality of distributor packages (Section 4.6). This labor suggests the need for distribution services to have professional personnel in the curation and moderation process. However, it is unclear to librarians whether content distributors have appropriate professional staff to handle curation tasks. This lack of clarity reduces librarians' confidence in what the distribution services offer. Therefore, services should reveal the qualifications of their content curation staff and describe the duties that this staff assumes to provide more insight into their curation practices. Additionally, if some crucial knowledge curation and content management tasks require additional personnel, they should hire qualified individuals to execute those tasks.

**5.4   Role of Other Stakeholders: Library Associations and Lawmakers**

Librarian associations such as the Public Library Association (PLA) or the American Library Association (ALA) could also help address digital curation challenges. They could publicize the problems among their constituent libraries and generate collective support for actions that could bring content distributors to the discussion table. They could also handle the distributor negotiations on behalf of participating libraries, giving them far more negotiating power in demanding changes from distributor services. As our analysis shows, inappropriate digital materials get reported less frequently, and many librarians may not be aware of them. Therefore, these associations' efforts to increase awareness about content deficiencies could help alert many public libraries to implement necessary remedial steps. These associations could also call for greater official recognition and formalization of the librarian tasks associated with handling digital content curation.

We argue that regulatory attention to digital curation and moderation problems introduced by the subscription model is crucial to protecting users' access to high-quality information and promoting libraries' public service goals. The various benefits this model offers public libraries (Section 4.1), including cost savings, suggest that it will continue to be adopted by the libraries, especially in the face of recent budget cuts [56]. Therefore, lawmakers might set minimum content quality thresholds that distributor services should include in their products and require regular audits to ascertain the services' compliance. Laws could also be instituted to incorporate greater transparency into digital distributors' content curation and moderation practices. Finally, lawmakers could incentivize more firms to offer digital distribution services to break the current marketplace monopoly (Section 4.2.3).

We found many similarities between the content management challenges of digital distribution services and social media platforms. Thus, laws regulating social media content may also serve as a valuable guide for creating new laws that shape digital content distributors' practices.

### 5.5   Limitations and Future Work

Our focus on public librarians allowed us to understand their mental models of how digital distributor services operate. While we attempted to arrange interviews with the distribution service staff through various channels, we received no response. Future investigations that resolve this challenge and surface the perspectives and practices of distribution service staff should offer essential complementary insights to our work.

Most study participants were White and had more than ten years of work experience. Their duties included handling digital content, and most participants oversaw adult content. Most worked in libraries in the US Midwest and East coast regions. These participant characteristics have shaped our findings. Recruiting a broader pool of librarians and employing complementary data collection methods, such as large-scale surveys, should generate additional valuable insights.

Some participants mentioned receiving more content review requests for physical materials than digital ones (Section 4.2.2). This difference raises questions about the discoverability of inappropriate digital content and how patrons view its inclusion in the library. Several participants also mentioned freedom of speech as an essential concern for their curation practices (Section 4.4.1). Excluding information that could otherwise have been accessed on the distributor website invites speculation that the library engages in censorship. Thus, moderation of library content requires librarians to avoid including biased content and prevent infringing on free speech values simultaneously [57]. Given the complexity of this challenge, future research on digital content moderation in library settings should consider how distribution services and librarians can cooperate to protect users' digital access to diverse materials and reduce harm from problematic information.

## 6   CONCLUSION

This paper examined the problem of inappropriate content available in public libraries through digital distribution services and the content curation and moderation deficiencies of such offerings. Thus far, this issue has surfaced only in scattered media articles [58], but its prominence and long-term effects on the broader public are only likely to grow as libraries become increasingly digital. We anticipate that a few major distribution companies will continue dominating the content provider landscape. Therefore, we must begin addressing the challenges our research identified and ask these corporations to improve their practices and work with librarians as equal partners. We hope that our empirical research would help build advocacy and support for such initiatives. Further, our integration of CSCW research on content moderation here suggests that scholarship in CSCW and library sciences can inform each other towards a better understanding of and solutions to digital governance challenges.

**REFERENCES**

[1] Audunson, R., Essmat, S. and Aabø, S. Public libraries: A meeting place for immigrant women? *Library & information Science Research*, 33, 3 (2011), 220-227

[2] Scott, R. The Role of Public Libraries in Community Building. *Public Library Quarterly*, 30, 3 (2011), 191-227 https://doi.org/10.1080/01616846.2011.599283.

[3] Vårheim, A. Trust and the role of the public library in the integration of refugees: The case of a Northern Norwegian city. *Journal of Librarianship and Information Science*, 46, 1 (2014), 62-69

[4] Hansson, J. *Libraries and identity: the role of institutional self-image and identity in the emergence of new types of libraries*. Elsevier, 2010.

[5] Stenstrom, C., Cole, N. and Hanson, R. A review exploring the facets of the value of public libraries. *Library Management*, 40, 6/7 (2019), 354-367 https://doi.org/10.1108/lm-08-2018-0068.

[6] Lor, P., Wiles, B. and Britz, J. Re-thinking information ethics: truth, conspiracy theories, and librarians in the COVID-19 era. *Libri*, 71, 1 (2021), 1-14

[7] Gorwa, R. What is platform governance? *Information, Communication & Society*, 22, 6 (2019), 854-871 https://doi.org/10.1080/1369118x.2019.1573914.

[8] Helberger, N., Pierson, J. and Poell, T. Governing online platforms: From contested to cooperative responsibility. *The Information Society*, 34, 1 (2017), 1-14 https://doi.org/10.1080/01972243.2017.1391913.

[9] Gillespie, T. *Custodians of the Internet: Platforms, content moderation, and the hidden decisions that shape social media*. Yale University Press, 2018.

[10] Geiger, A. w. 2017. *Most Americans – especially Millennials – say libraries can help them find reliable, trustworthy information*. 2023 Feb 1. https://www.pewresearch.org/fact-tank/2017/08/30/most-americans-especially-millennials-say-libraries-can-help-them-find-reliable-trustworthy-information/

[11] Grimmelmann, J. The Virtues of Moderation. *Yale Journal of Law & Technology*, 42 (2015), 44-109

[12] Hawthorne, D. *History of electronic resources*. IGI Global, City, 2008.

[13] Morris, C. and Sibert, L. *Acquiring E-books*. American Library Association, 2010.

[14] Schroeder, R. Doing More with Less Adoption of a Comprehensive E-book Acquisition Strategy to Increase Return on Investment while Containing Costs. *LRTS*, 62, 1 (2018), 28-36

[15] Platt, C. Popular E-content at the New York public library: Successes and challenges. *Publishing Research Quarterly*, 27, 3 (2011), 247-253 https://doi.org/10.1007/s12109-011-9231-6.

[16] Walters, W. H. E-books in Academic Libraries: Challenges for Acquisition and Collection Management. *Portal: Libraries and the Academy*, 13, 2 (2013), 187-211 https://doi.org/10.1353/pla.2013.0012.

[17] Waugh, M., Donlin, M. and Braunstein, S. Next-Generation Collection Management: A Case Study of Quality Control and Weeding E-Books in an Academic Library. *Collection Management*, 40, 1 (2015), 17-26 https://doi.org/10.1080/01462679.2014.965864.

[18] Vaidhyanathan, S. Afterword: Critical Information Studies. *Cultural Studies*, 20, 2-3 (2006), 292-315

[19] Oliphant, T. A case for critical data studies in Library and Information Studies. *Journal of Critical Library and Information Studies*, 1, 1 (2017)

[20] Bossaller, J., Adkins, D. and Thompson, K. M. Critical theory, libraries and culture. *Progressive Librarian*, 34/35 (2010), 25

[21] Barreau, D. Information systems and collection development in public libraries. *Library Collections, Acquisitions, & Technical Services*, 25, 263–279 (2001)

[22] Sullivan, M. Libraries and fake news: What's the problem? What's the plan? *Communications in Information Literacy* (2019)
[23] De Paor, S. and Heravi, B. Information literacy and fake news: How the field of librarianship can help combat the epidemic of fake news. *The Journal of Academic Librarianship*, 46, 5 (2020), 102218
[24] Audunson, R. The public library as a meeting-place in a multicultural and digital context: The necessity of low-intensive meeting-places. *Journal of Documentation* (2005)
[25] Langvardt, K. Regulating online content moderation. *Georgetown Law Journal*, 106 (2017)
[26] Jhaver, S., Birman, I., Gilbert, E. and Bruckman, A. Human-Machine Collaboration for Content Regulation. *ACM Transactions on Computer-Human Interaction*, 26, 5 (2019), 1-35 https://doi.org/10.1145/3338243.
[27] Kiesler, S., Kraut, R., Resnick, P. and Kittur, A. *Building successful online communities: Evidence-based social design.* MIT Press, City, 2010.
[28] Seering, J., Kraut, R. and Dabbish, L. Shaping Pro and Anti-Social Behavior on Twitch Through Moderation and Example-Setting. In *Proceedings of the Proceedings of the 2017 ACM Conference on Computer Supported Cooperative Work and Social Computing* (2017).
[29] Carlson, C. R. and Cousineau, L. S. Are you sure you want to view this community? exploring the ethics of reddit's quarantine practice. *Journal of Media Ethics*, 35, 4 (2020), 202-213
[30] Myers West, S. Censored, suspended, shadowbanned: User interpretations of content moderation on social media platforms. *New Media & Society*, 20, 11 (2018), 4366-4383 https://doi.org/10.1177/1461444818773059.
[31] Seering, J., Wang, T., Yoon, J. and Kaufman, G. Moderator engagement and community development in the age of algorithms. *New Media & Society*, 21, 7 (2019), 1417-1443 https://doi.org/10.1177/1461444818821316.
[32] Singhal, M., Ling, C., Stringhini, N. K. G. and Nilizadeh, S. SoK: Content Moderation in Social Media, from Guidelines to Enforcement. *ArXiv Preprint* (2022)
[33] Fiesler, C., McCann, J., Frye, K. and Brubaker, J. R. Reddit rules! characterizing an ecosystem of governance. In *Proceedings of the Twelfth International AAAI Conference on Web and Social Media.* (2018).
[34] Jiang, J. A., Middler, S., Brubaker, J. R. and Fiesler, C. Characterizing Community Guidelines on Social Media Platforms. In *Proceedings of the Conference Companion Publication of the 2020 on Computer Supported Cooperative Work and Social Computing* (2020).
[35] Chandrasekharan, E., Gandhi, C., Mustelier, M. W. and Gilbert, E. Crossmod: A Cross-Community Learning-based System to Assist Reddit Moderators. *Proceedings of the ACM on Human-Computer Interaction*, 3, CSCW (2019), 1-30 https://doi.org/10.1145/3359276.
[36] Banchik, A. V. Disappearing acts: Content moderation and emergent practices to preserve at-risk human rights–related content. *New Media & Society*, 23, 6 (2020), 1527-1544 https://doi.org/10.1177/1461444820912724.
[37] Gorwa, R., Binns, R. and Katzenbach, C. Algorithmic content moderation: Technical and political challenges in the automation of platform governance. *Big Data & Society*, 7, 1 (2020) https://doi.org/10.1177/2053951719897945.
[38] He, Q., Hong, Y. K. and Raghu, T. S. The Effects of Machine-powered Content Moderation: An Empirical Study on Reddit. In *Proceedings of the In 55th Hawaii International Conference on System Sciences (HICSS).* (Hawaii, 2022).
[39] Jhaver, S., Appling, D. S., Gilbert, E. and Bruckman, A. "Did You Suspect the Post Would be Removed?". *Proceedings of the ACM on Human-Computer Interaction*, 3, CSCW (2019), 1-33 https://doi.org/10.1145/3359294.

[40] Juneja, P., Rama Subramanian, D. and Mitra, T. Through the Looking Glass. *Proceedings of the ACM on Human-Computer Interaction*, 4 (2020), 1-35 https://doi.org/10.1145/3375197.
[41] Jhaver, S., Bruckman, A. and Gilbert, E. Does Transparency in Moderation Really Matter? *Proceedings of the ACM on Human-Computer Interaction*, 3, CSCW (2019), 1-27 https://doi.org/10.1145/3359252.
[42] Suzor, N. P., West, S. M., Quodling, A. and York, J. What do we mean when we talk about transparency? toward meaningful transparency in commercial content moderation. *International Journal of Communication*, 13, 18 (2019), 1526–1543
[43] McInnis, B., Ajmani, L., Sun, L., Hou, Y., Zeng, Z. and Dow, S. P. Reporting the Community Beat: Practices for Moderating Online Discussion at a News Website. *Proceedings of the ACM on Human-Computer Interaction*, 5, CSCW2 (2021), 1-25
[44] Pater, J. A., Kim, M. K., Mynatt, E. D. and Fiesler, C. Characterizations of Online Harassment. In *Proceedings of the Proceedings of the 19th International Conference on Supporting Group Work* (2016).
[45] Chandrasekharan, E., Samory, M., Jhaver, S., Charvat, H., Bruckman, A., Lampe, C., Eisenstein, J. and Gilbert, E. The Internet's Hidden Rules. *Proceedings of the ACM on Human-Computer Interaction*, 2, CSCW (2018), 1-25 https://doi.org/10.1145/3274301.
[46] Crawford, K. and Gillespie, T. What is a flag for? Social media reporting tools and the vocabulary of complaint. *New Media & Society*, 18, 3 (2014), 410-428 https://doi.org/10.1177/1461444814543163.
[47] Seering, J. Reconsidering Self-Moderation. *Proceedings of the ACM on Human-Computer Interaction*, 4, CSCW (2020), 1-28 https://doi.org/10.1145/3415178.
[48] DeNardis, L. and Hackl, A. M. Internet governance by social media platforms. *Telecommunications Policy*, 39, 9 (2015), 761-770 https://doi.org/10.1016/j.telpol.2015.04.003.
[49] Kiene, C., Monroy-Hernández, A. and Hill, B. M. Surviving an" Eternal September" How an Online Community Managed a Surge of Newcomers. In *Proceedings of the Proceedings of the 2016 CHI Conference on Human Factors in Computing Systems* (2016).
[50] Merriam, S. B. Introduction to qualitative research. *Qualitative Research in Practice: Examples for Discussion and Analysis*, 1, 1 (2002), 1-17
[51] Paris, B., Carmien, K. and Marshall, M. "We want to do more, but…": New Jersey public library approaches to misinformation. *Library & Information Science Research*, 44, 2 (2022), 101157
[52] Hoopla. 2015. *What is hoopla's content policy?* 2023 Feb 1. https://www.hoopladigital.com/help
[53] Tunick, M. *Balancing privacy and free speech: Unwanted attention in the age of social media*. Taylor & Francis, 2015.
[54] Haimson, O. L., Delmonaco, D., Nie, P. and Wegner, A. Disproportionate removals and differing content moderation experiences for conservative, transgender, and black social media users: Marginalization and moderation gray areas. *Proceedings of the ACM on Human-Computer Interaction*, 5, CSCW2 (2021), 1-35
[55] Jhaver, S., Chen, Q. Z., Knauss, D. and Zhang, A. X. *Designing Word Filter Tools for Creator-led Comment Moderation*. City, 2022.
[56] Association, A. L. 2018. *Budget in the Crosshairs? Navigating a Challenging Budget Year*. 2023 Feb 1. https://www.ala.org/advocacy/navigating-challenging-budget-year-budget-crosshairs
[57] Walsh, J. Librarians and controlling disinformation: is multi-literacy instruction the answer? *Library Review*, 59, 7 (2010), 498-511

[58] Woodcock, C. 2022. *Ebook Services Are Bringing Unhinged Conspiracy Books into Public Libraries*. Feb 1. https://www.vice.com/en/article/93b7je/ebook-services-are-bringing-unhinged-conspiracy-books-into-public-libraries